# Scaling While Privacy Preserving: A Comprehensive Synthetic Tabular Data Generation and Evaluation in Learning Analytics

Qinyi Liu

Centre for the Science of Learning & Technology (SLATE), University of Bergen, qinyi.liu@uib.no

Mohammad Khalil

Centre for the Science of Learning & Technology (SLATE), University of Bergen, mohammad.khalil@uib.no

Ronas Shakya

Centre for the Science of Learning & Technology (SLATE), University of Bergen, ronas.shakya@student.uib.no

Jelena Jovanovic

Faculty of Organisational Sciences, University of Belgrade; Centre for the Science of Learning & Technology (SLATE), University of Bergen, jelena.jovanovic@uib.no

Privacy poses a significant obstacle to the progress of learning analytics (LA), presenting challenges like inadequate anonymization and data misuse that current solutions struggle to address. Synthetic data emerges as a potential remedy, offering robust privacy protection. However, prior LA research on synthetic data lacks thorough evaluation, essential for assessing the delicate balance between privacy and data utility. Synthetic data must not only enhance privacy but also remain practical for data analytics. Moreover, diverse LA scenarios come with varying privacy and utility needs, making the selection of an appropriate synthetic data approach a pressing challenge. To address these gaps, we propose a comprehensive evaluation of synthetic data, which encompasses three dimensions of synthetic data quality, namely resemblance, utility, and privacy. We apply this evaluation to three distinct LA datasets, using three different synthetic data generation methods. Our results show that synthetic data can maintain similar utility (i.e., predictive performance) as real data, while preserving privacy. Furthermore, considering different privacy and data utility requirements in different LA scenarios, we make customized recommendations for synthetic data generation. This paper not only presents a comprehensive evaluation of synthetic data but also illustrates its potential in mitigating privacy concerns within the field of LA, thus contributing to a wider application of synthetic data in LA and promoting a better practice for open science.

# 1 INTRODUCTION

A prominent concern that has been present in the learning analytics (LA) community since the inception of the field is how to replicate LA solutions at scale. This scaling entails, for example, the breadth of LA applications, the transition from ad-hoc to post-hoc solutions, and the expansion from specialization to generalization [34, 56]. Privacy in a congruent way, has also become a significant barrier to scaling the field [15]. In their recent review, [32] outlined eight distinct privacy and data protection issues in phases of the LA process. Concerns about privacy issues continue to escalate with the increasing volume of data, the diversity of data patterns, the regular updates of data protection laws, and the popularity of generative artificial intelligence [23]. In contrast, current approaches to addressing privacy issues in LA have some important limitations. For example, legal- and framework-based solutions often lack practical evidence and clear and actionable guidelines [32, 36]. Approaches, such as the Privacy and Data Protection Framework [46], point towards practical challenges in implementing privacy measures in the field of LA, but do not suggest how to overcome them. Furthermore, [54] proved that current technical de-identification measures were vulnerable to attacks based on unsupervised machine learning methods. Such a lack of proper anonymization creates opportunities for data misuse. Regulations of the local storage and processing of sensitive data, such as the General Data Protection and Regulation act (GDPR) [29] are also difficult to fully address with current privacy solutions in LA [32]. In this context, synthetic data holds significant promise for mitigating privacy concerns in the field of LA and for expanding analytics capabilities without exposing individual details [5].

Synthetic data refers to data generated using mathematical models [24]. It can serve multiple functions in various domains. For example, it can be used to substitute hard-to-obtain samples and reduce the data collection cost [49]. In terms of privacy protection, synthetic data can fulfill the requirement of data protection regulations by serving as a substitute for sensitive data that cannot be migrated. Moreover, synthetic data can be used as a substitute for the original data in situations where data sharing is required, to protect the privacy of the original data, thus addressing insufficient anonymization and data misuse from another perspective. [25] argued that synthetic data would not be considered identifiable personal data, thus privacy regulations would not apply. In general, synthetic data can be evaluated from three perspectives: i) structural similarity to the original data (resemblance), ii) its usage for analysis and predictive modeling (utility), and iii) protection of information stored in the real data (privacy) [24]. Regarding the application domains, synthetic data has been used in a variety of domains, e.g., in health [55], in robotics training [37], and in transportation sign detection [31]. In the field of LA, there have been only a few applications of synthetic data, mainly in the form of synthetic tabular or sequence data generation (see Section 2.3). However, these studies also have certain limitations. A notable one is that when evaluating the quality of synthetic data, they tend to consider only some of the aforementioned dimensions (i.e., resemblance, utility, and privacy), and thus offer a narrow perspective on the synthetic data quality. For example, [4] only evaluate resemblance, lacking privacy and utility, whereas [57] evaluate utility and resemblance, without evaluating privacy.

Generally, previous synthetic data-related research in the field of LA has not provided a comprehensive and multidimensional evaluation of synthetic data. The lack of a comprehensive evaluation makes it difficult to assess the suitability of synthetic data for particular LA usage scenarios, thus hampering wider use of synthetic data in the field of LA and preventing the field from benefiting from the many advantages of synthetic data. Moreover, the previous attempts to generate synthetic data in the LA domain were not informed by the actual use of such data, that is, specific scenarios in which the generated synthetic data would be used were not considered. For example, in some cases, the use of synthetic data is due to the pursuit of higher privacy protection, whereas in some other cases, synthetic data is used for the sake of achieving higher data utility (e.g., higher performance of a model predicting at-risk students) and it is possible to trade-off the improvement in privacy for higher data utility. How to make the synthetic data aligned with the needs of different usage scenarios in LA research and practice is still an open question. To contribute to bridging this gap, the current study focuses on synthetic tabular data and address the following research questions:

RQ1: How to make a comprehensive assessment of synthetic tabular data, including the dimensions of resemblance, utility, and privacy, in the LA field?
RQ2: To what extent the use of synthetic tabular data can improve the privacy aspects in LA predictive modeling while maintaining the model's predictive performance?
RQ3: How to customize the generation of synthetic tabular data to suit different scenarios in LA predictive modeling?

The contributions of our work are as follows:

- To our knowledge, this paper provides a comprehensive, three-dimensional assessment of synthetic tabular data (i.e., resemblance, utility, and privacy) for the first time in the LA field. By enabling LA researchers to make well informed decisions regarding the synthetic data use, such a comprehensive three-dimensional evaluation would help to allay concerns and promote a wider use of synthetic data in the LA field.
- We show that synthetic data can maintain similar predictive performance as real data while being privacy-enhanced. This demonstrates that synthetic data can be applied as a more reliable privacy-preserving means in the LA domain.
- We discuss the need for privacy and predictive modeling in different scenarios of LA, and how to customize the synthetic data to better fit the specific requirements.

## 2 RELATED WORK

### 2.1 Privacy in learning analytics

LA has come a long way since its inception in 2011. However, privacy concerns have also been on a steady rise and have been identified as a major obstacle to the further development of the field [15, 32]. Privacy issues differ at distinct stages of a LA project. In the phases of data analytics and data publication and sharing, insufficient anonymization, sensitive data storage, and calculation are urgent privacy issues [32]. Current de-identification and anonymization techniques may not provide the necessary privacy protection. For example, [54] found that they could use unsupervised learning to rediscover information about students from de-identified data. [21] further note that due to the volume and abundance of data from online courses in recent years, a variety of previously unavailable identifiers continue to populate the database, and existing means of anonymization are not sufficient. Sensitive data storage and calculation is an issue that has emerged alongside rules, such as GDPR [29], which have imposed restrictions on the storage, migration, and computation of data containing sensitive information for security and privacy reasons. Such initiatives will undoubtedly place additional restrictions on data analysis and data sharing in LA. In the phase of data publication and sharing, in addition to the previously mentioned privacy issues, another potential privacy issue is data misuse [32]. When LA data is released for public use, the rights of students can be violated if someone tries to use it inappropriately. A prime example of this is the discovery by [30] that student data collected by the Early Alert Student Indicator (EASI) at the University of Queensland, Australia, was used in ways different than approved by the student consent.

With the increasing awareness of the distinct privacy issues [32, 33] and their impact on the further scaling of LA, initial solutions started to emerge in the LA community. However, few are directly beneficial to solving the privacy issues in LA, namely anonymization, sensitive data storage and calculation, and data misuse. Notable examples include two MOOC-based privacy-preserving tools developed by [47] and [21]. The former is a browser-based privacy-preserving data preprocessing tool that allows researchers to store sensitive information locally [47]. The latter is a tool that allows LA data to be used in a "usable invisible" way, i.e., researchers can use the data but they cannot see the data [21]. However, these approaches still have certain limitations. First, both tools can only be applied to MOOC scenarios. Second, neither method actually share or publish the data for the public to use, which ensures privacy but also creates an obstacle for further development of open science. Another work related to anonymization is the privacy risk quantification with Markov model, proposed by [50], which can be used to assess the re-identification risk of educational tabular data before it is shared. However, the method is complex and has certain technical barriers for use. Additionally, this model does not consider the next step after measuring the risk. If the re-identification risk is high, increase in anonymization will still result in potential trade-off between privacy and utility.

In addition to the aforementioned technical solutions that directly address data privacy issues, there are also legal and framework-based solutions. For example, [26] proposed a conceptual framework to de-identify learning analytics data. Another example is by [42] who suggest that appropriate ethical behavior must be enforced through contracts to effectively address privacy issues in LA. While these legal and framework-based solutions can provide some guidance in terms of overall LA design, they tend to be conceptual and lack clear actionable guidelines and evidence of application (as discussed by [32, 36].

As a novel avenue for addressing privacy issues in LA, synthetic data generation has been recently proposed in computer science [24]. Synthetic data generation may have potential to balance data utility and privacy, be generalizable to different types and sources of LA data and facilitate data publication and sharing for open science [5, 57].

### 2.2 Synthetic data generation and evaluation

According to [24], synthetic data is "data that has been generated using a purpose-built mathematical model or algorithm, with the aim of solving a (set of) data science task(s)" (p.5). In terms of functions, synthetic data has three main functions. The first is to serve as a substitute for data that cannot be released to the outside world for privacy reasons, thus supporting open science and accessible data quality. Second, bias in machine learning models can be reduced by increasing the proportion of minority samples, as well as simulating what-if scenarios through synthetic data. Third, the expense of data labeling can be reduced by generating synthetic data [24]. Additionally, there are three types of synthetic data, namely text data, tabular data, and media data [49]. In this paper, we focus on synthetic tabular data. Tabular data can be defined as data that is structured in a tabular form [22]. It does not inherently imply a sequential order of rows, and the rows are typically independent of each other. Relevant to our work, the following subsections describe the generation and evaluation of synthetic data, as well as its use in LA.

*2.2.1 Synthetic data generation*

The most common synthetic tabular data generation methods can be divided into two main types, namely statistical distribution methods and deep learning-based methods [28, 11]. The principle of the former group is to generate appropriate samples by considering the probability distribution of the real dataset. Representative methods include, for example, Gaussian multivariate [16], Chi-square, and Monte Carlo method [28]. While statistical distribution methods offer advantages such as speed, ease of application, minimal computing resource demands, and typically manageable parameters, they may not be suitable for handling large or complex data [28, 20]. Deep learning, particularly generative adversarial networks (GANs), has rapidly gained traction due to its computation efficiency [20]. There are many variants of GAN-based approaches, such as Wasserstein GAN [2] for hyperparametric search, HealthGAN [55] for health domain, and conditional GAN (CTGAN) for synthetic tabular data [53].

Furthermore, there is an important ecosystem for synthetic tabular and time series data generation, namely Synthetic Data Vault (SDV) [41]. SDV consists of a combination of multiple probabilistic graphical modeling and deep learning (DL) based techniques [20]. SDV is widely used in different fields of synthetic generation and has been shown to achieve similar results as using real data in the tested datasets [41, 58].

In this paper, we use one method from each of the three aforementioned groups: i) Gaussian Multivariate from statistical distribution methods, which is suitable for small datasets and is computationally lightweight; ii) CTGAN from the group of deep learning-based methods, which perform well in tabular datasets with complex relationships, and iii) GaussianCopula model from SDV, as it can reconstruct relationships between real data variables better and perform well on LA data [57]. The details of three methods are presented in the methodology section (Section 3.2).

*2.2.2 Synthetic data evaluation*

Synthetic data must closely resemble the original (real) data and replicate its crucial statistical characteristics. However, it should not mirror the data too closely, as this can compromise privacy. Synthetic data should also possess a certain level of utility to perform '*as well as possible*' in data analytics tasks, such as machine learning predictive models. Hence, researchers commonly assess synthetic data based on three attributes [24]: 1) resemblance, 2) utility, and 3) privacy.

**Resemblance**. The resemblance, or commonly known as fidelity or similarity [24, 58], refers to resembling synthetic data to real data. [10] have provided a detailed categorization of resemblance metrics, including univariate fidelity, bivariate fidelity, population fidelity, and application fidelity. Univariate fidelity requires that each attribute in the synthetic data has a similar structure of basic summary statistics as the corresponding attribute in the real data [10]. The representative methods include Hellinger distance [35] and Kullback-Leibler divergence [7]. Whereas bivariate fidelity builds on univariate fidelity, it refers to a measure of correlation between pairs of variables in a dataset to capture the statistical dependence structure of the real and synthetic datasets [10]. A representative approach is pairwise correlation [9]. Population fidelity (also known as multivariate fidelity), on the other hand, reflects large-scale features of the entire distributions of the synthetic dataset in comparison with the real dataset. It is currently the most popular method [9]. Commonly used multivariate fidelity methods include Kolmogorov-Smirnov type statistics and Distinguishability type metrics, to name but a few [10]. Finally, application fidelity refers to prediction accuracy, which overlaps with the utility dimension [24].

In this paper, we use the Jensen-Shannon divergence (JSD), as a representative of the univariate fidelity measures. JSD is a symmetric version of the Kullback-Leibler divergence (KLD), and compared to KLD, JSD is always finite [9]. It is widely used for evaluating synthetic data generation based on deep learning methods. Additionally, we use the Wasserstein distance (WD) from population fidelity because it can better address the difficulties of KLD and Total Variation distance on discontinuous mappings [6]. For the bivariate fidelity measure of correlation between pairs of variables, this paper uses the well-recognized method of pairwise correlation [9]. The details of these resemblance metrics are presented in the methodology section (Section 3.3.2).

**Utility**. The evaluation of synthetic data on the utility dimension is mainly done on machine learning models [20, 10]. The train-on-real test-on-real (TRTR) and train-on-synthetic test-on-real (TSTR) method is recognized as effective and widely used for evaluating the utility of synthetic data [20]. The gist of this method is training the machine learning model on real and synthetic data separately and using the held-out real dataset for prediction. The two models are compared using performance measures such as Accuracy and F1 score, to measure the utility of the synthetic data [20]. Such an approach is used in this paper, and the details of its use are presented in the methodology section (Section 3.3.1).

**Privacy**. Privacy measures can be divided into two categories, namely distance and similarity metrics and re-identification risk assessment [20]. The former overlaps partially with the resemblance. The reason behind this is straightforward: if the synthetic data is too similar to the original data, there is a high privacy risk. Among distance and similarity metrics, distance to closest record (DCR) and nearest neighbor distance ratio (NNDR) are commonly used to measure the privacy of synthetic tabular data [58, 45]. DCR and NNDR have also been applied at the industrial level [48]. Furthermore, DCR and NNDR can provide quantitative and easily interpretable results. For all these reasons, DCR and NNDR were selected for the current study.

As for the re-identification risk assessment, it refers to assessing the re-identification of real data disclosure risk through synthetic data generation [30]. One of the most used methods is membership interface attack (MIA). MIA in a synthetic data context means that an attacker tries to identify if real records have been used to train the synthetic data generation algorithm [30]. The second most used method is attribute interface attack (AIA), which is an attempt to take a known portion of the attributes of the real data and guess the value of an unknown attribute of the synthetic data [30]. In this paper, only MIA is used. AIA requires attributes considered quasi-identifiers, which are not present in all datasets used in our study.

### 2.3 Synthetic data in LA

The role of synthetic data in LA was scarcely explored. One of the earliest studies is by [5] who argued that the LA community should adopt synthetic data generators. [5] also argued that synthetic data could meet the urgent demand for richer datasets as well as the need for safeguarding students' privacy. Later, [39] explored the use of synthetic tabular data in LA. Based on their experimental results, they concluded that synthetic data may be used for prediction and classification tasks in LA and called on exploring synthetic data to improve data accessibility and collaboration. Although the two initiatives by [5] and [39] have shown merit, they entailed limitations, primarily related to non-experimental research or small size studies. Following this, [4] explored if synthetic sequence and tabular data could serve as a privacy-preserving data source. They used the least squares Generative Adversarial Network (LSGAN) method to create similar, but slightly different, privacy-preserving synthetic student data. However, the method by [4] underperforms algorithmically when faced with data of multiple features. Furthermore, [4] evaluated only the resemblance aspect of the synthetic data, without evaluating the utility and privacy aspects. [51], on the other hand, used synthetic educational data to show that either the naïve anonymization or removing rows from a dataset was not enough for de-identification. [51] also presented an evaluation framework for comparing synthetic data generators. However, the framework is limited, with evaluations of the utility and privacy aspects only, and the evaluation of utility restricted to weighted Root Mean Square Error (RMSE) only. [57] focused on the performance comparison of synthetic tabular data generators in the context of LA predictive models. They evaluated and tested four different synthetic data generators and showed that synthetic data generators could replicate the statistical structure of real data and exhibit better machine learning performance. While their work focuses on the evaluation of utility and resemblance, it lacks the further consideration of privacy.

## 3 METHODOLOGY

Figure 1 illustrates the method we propose for a comprehensive assessment of synthetic tabular data in the LA domain (RQ1). The figure depicts the three key phases of the method: (original) data preprocessing (Section 3.1), synthetic data generation (Section 3.2), and synthetic data evaluation (Section 3.3). In addition to addressing RQ1, the presented method also allows for addressing RQ2. Section 3.4 presents how RQ3 is addressed.

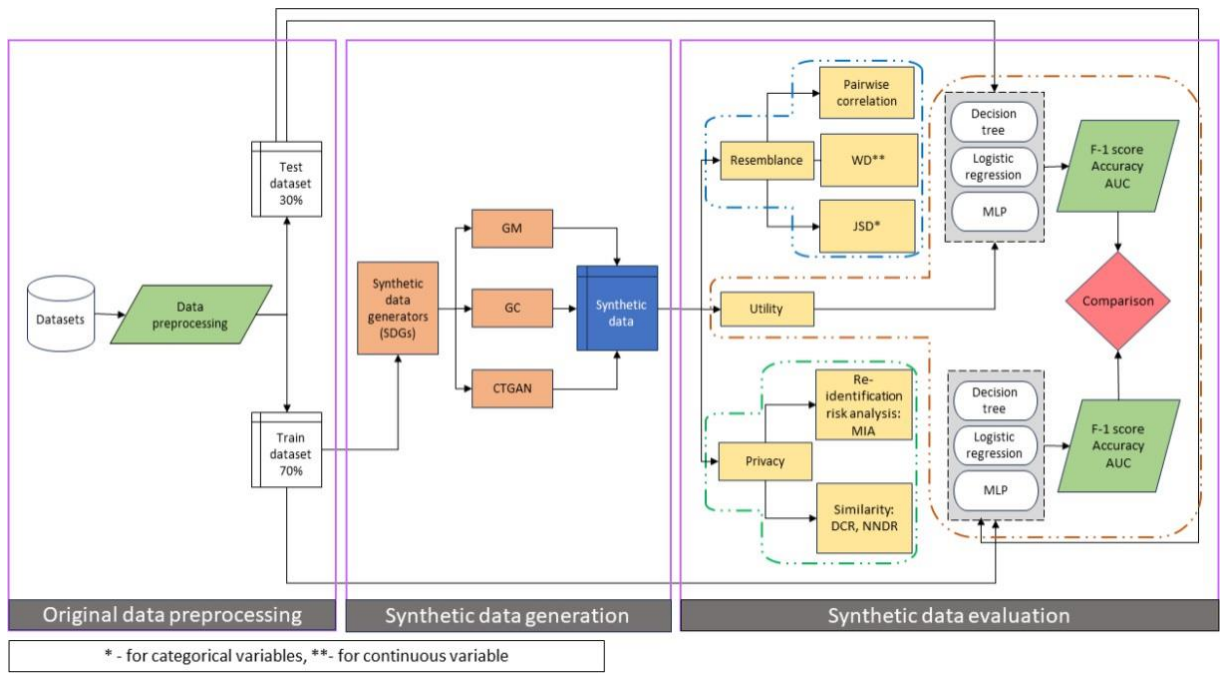

Figure 1. The workflow of the proposed method for comprehensive assessment of synthetic tabular data in LA

### 3.1 Dataset selection and pre-processing

We chose three LA datasets with different original data volume to model different scenarios. We further increased the representativeness by choosing differently balanced datasets, as suggested by [19]. A brief description of the datasets, including the number of attributes and records, is presented in Table 1. Two of the datasets (A and B) are from publicly available datasets repository on Kaggle, whereas the third dataset (C) is not publicly available at this time. The three datasets notably differ in terms of size, with the number of records equal to or below 1000, 500, and 100, respectively. In terms of balance, as denoted by the imbalance ratio (Mi) in Table 1, dataset A is not as balanced as datasets B and C.

Each dataset was pre-processed, which included filtering out missing values[1] and splitting it into two subsets: 70% of the records were used as the input for the synthetic tabular data generation methods (Section 3.2), whereas the remaining 30% of the records were used for evaluation (Section 3.3).

Table 1: Brief description of the selected LA datasets (before pre-processing). The notations $C$, $B$, and $M$ represent the number of continuous, binary, and multi-class categorical variables, respectively, whereas $Mi$ stands for the imbalance ratio (i.e., the ratio of minority and majority samples).

| ID | Dataset name | Year | Num. attrib. | Num. records | Target variable | #C | #B | #M | #Mi |
|----|---|---|---|---|---|---|---|---|---|
| A | Students Performance in Exams* | 2018 | 8 | 1000 | Continuous | 3 | 3 | 2 | 0.05 |
| B | Students' Academic Performance** | 2016 | 17 | 480 | Multi-class | 4 | 4 | 9 | 0.67 |
| C | Private MOOC dataset | 2021 | 12 | 64 | Continuous | 12 | 0 | 0 | 0.64 |

* Dataset retrieved from (https://www.kaggle.com/datasets/spscientist/students-performance-in-exams), CC-BY licensed
** Dataset retrieved from (https://www.kaggle.com/datasets/aljarah/xAPI-Edu-Data ) licensed CC-BY-SA 4.0

### 3.2 Synthetic tabular data generation

We used three open-source data generative methods: (1) Gaussian Multivariate (GM) by [16], (2) Gaussian Copula by [41], and (3) Conditional Tabular Generative Adversarial Network (CTGAN) by [53]. Brief descriptions of these methods are given below, while the rationale for their selection is outlined in Section 2.2.1.

*Gaussian Multivariate* (GM) is one of the classic approaches for generating tabular data, which implements multivariate distribution to combine marginal probabilities estimated from univariate distributions [16]. The advantage of GM is that it supports several multivariate distributions, allowing us to work with multiple random variables while considering the dependencies that may exist between them (Multivariate Distributions — Copulas 0.9.1 Documentation, 2021)

*Gaussian Copula (GC)* is a statistical model to measure the dependence structures between variables by comparing the joint distribution of the variables and their marginal distributions. The advantage of GC is that it can model both covariances between original columns and conditional parameter dependencies on the original columns, making it a versatile choice for capturing complex multivariate dependencies in data [41]. In our study, we used the GC implementation from the SDV toolkit [41] and moved forward with recommended default parameters.

*Conditional Tabular Generative Adversarial Network (*CTGAN) is a GAN-based methodology, specialized in modeling tabular distributions and extracting row samples from these distributions. Its distinct advantages encompass the capability to mitigate training data imbalances via conditional generators and sampling training, as well as facilitating the generation of data with specific discrete values [53]. Moreover, tailored for tabular synthetic data generation, CTGAN exhibits high performance in tabular data applications, outperforming most of Bayesian and deep learning methodologies [53]. In our study, we harnessed CTGAN 0.7.4 from the SDV toolkit [41], adhering to its recommended default parameters.

### 3.3 Synthetic tabular data evaluation

To examine and compare the performance of different synthetic data generation methods (Section 3.2), we compute distinct evaluation measures specific to each of the three dimensions of synthetic data quality (i.e., resemblance, utility and privacy)

---
[1] As a part of data pre-processing, small number of missing values (6) were removed from dataset C, only

on each of the datasets (Table 1). To enable an overall comparison of the performance of each synthetic data generator along these three dimensions, we used the method of [58] and [10], of calculating the average across the datasets. Formal definitions of the used evaluation metrics are available in the online supplementary document[2].

*3.3.1 Machine learning utility evaluation*

To evaluate the utility of the synthetic data, we compared the performance of predictive models built using the original and the synthetic data. As shown in Figure 1, for evaluation purposes we selected three classification algorithms widely used in LA: random forest classifier, (multinomial) logistic regression, and multi-layer-perceptron (MLP) [12, 14].

Seventy percent of the original data was used to train the three data generation models. Once the training was finished, we used the data generation models to generate synthetic tabular data of 5000 rows. The rationale for this size of the synthetic datasets is twofold: i) we aimed at a computationally feasible data set size, which would allow us to examine the effect of different ratios of real to synthetic data on the evaluation measures, ii) synthetic dataset of such size was used in an earlier similar study [18]. The synthetic dataset and the original training dataset were then separately used to train the above-mentioned three machine learning algorithms and the resulting models were evaluated on the original test dataset. The models' performance was measured via the Accuracy, F1-score, and Area under the ROC curve (ROC AUC).

*3.3.2 Resemblance evaluation*

To evaluate the synthetic data in terms of its resemblance to the original data, the following metrics were used:

**Difference in pairwise correlation**. Pairwise correlation is a measure of correlation between two expressions and is used to measure the difference between the results of two expressions [7]. Calculating the difference in pairwise correlations can effectively assess the degree of retention of feature interactions in a dataset, and for this purpose, we adopted the approach by [58]. We first computed the pairwise correlation matrices of the columns in the real and synthetic datasets. Pearson correlation coefficient and the Theil uncertainty coefficient were used to measure the correlation between continuous features and between categorical features, respectively [58]. Finally, the difference between the pairwise correlation matrices of the real and synthetic datasets was calculated [58]. The *dython* library is used to computed pairwise correlation.

**Jensen-Shannon divergence (JSD)** is a probabilistic and statistical measure of similarity between two probability distributions [38]. JSD has been used to measure the difference in the probability mass distribution of individual categorical variables between real and synthetic datasets [24]. As a symmetrical metric, with the [0, 1] range, JSD is easy to quantify and interpret, with a larger value indicating a greater gap between the two distributions, and vice versa [8].

**Wasserstein distance (WD)** is a method commonly used in the field of machine learning to compare probability distributions [10]. WD is used by a wide range of synthetic data generation methods, the classical Wasserstein GAN is an example [2]. WD can be used to measure how well a synthetic dataset corresponds to the real data when distributed over a single continuous/mixed variable and can compensate for the numerical instability of JSD when assessing the quality of continuous variables [58]. A smaller WD value indicates two more similar distributions, while a larger WD indicates a larger difference between the two [8]. Both JSD and WD were computed in python, based on their formulas, as given in the literature.

*3.3.3 Privacy evaluation*

To assess the privacy aspect of the generated synthetic data, we used the following metrics:

**Distance to Closest Record (DCR)** is the Euclidean distance between the record, say *r,* of the synthetic data and the *r's* nearest record in the original table [40]. DCR equal to zero means that the synthetic record will leak the real information, while higher DCR values mean less risk of privacy leakage. We also report DCR within Real Data and DCR within Synthetic Data, as these measures can help us to capture the model collapse problem that might arise [44]. The former metric is the Euclidean distance between each record in the real data and its closest corresponding record in the real data, whereas the latter measures the same but on the synthetic dataset.

**Nearest Neighbour Distance Ratio (NNDR)** is a classical empirical privacy evaluation method that measures the ratio of the Euclidean distance between the nearest and second nearest real neighbors and any corresponding synthetic record [24, 58]. NNDR is distributed in the [0, 1] range, with higher values indicating higher privacy. As for NNDR, we also report NNDR within Real Data and NNDR within Synthetic Data. Additionally, both NNDR and DCR are computed using a python script that implements these measures based on their formulas, as given in the literature.

---
[2] https://github.com/ql909/mathematical_definitions

**Membership interface attack (MIA)**. We adopt the MIA method used by [20]. As for the inner logic of MIA, it assumes that the attacker has access to all the records of the synthetic data and a randomly distributed subset of the real data. The attacker will use the student records ($r$) in the real data subset and try to find the closest record in the synthetic data by calculating the distance metric. For computing this privacy measure, we used the script developed by [20][3].

### 3.4 Customizing synthetic tabular data generation to different scenarios

To address RQ3, we leveraged the weighting score approach for assessing the overall performance of synthetic tabular data, proposed by [20]. Three different scenarios are examined: i) relatively balanced demand for resemblance, utility, and privacy, where the performance of each dimension is equally weighted, such as the case of an independent LA researcher who wants to equally cover all dimensions; ii) high demand for utility and resemblance but little demand for privacy, such as in the context of an internal audit in an educational institution, and iii) low demand for utility but high demand for privacy, such as the case when a LA dataset is shared with an external technology company to validate synthetic data generation. Based on the results of the evaluation metrics (Section 3.3), we selected the algorithms and designs corresponding to the given scenarios and recommended the optimal configuration for the generation of synthetic tabular data for each LA scenario.

## 4 FINDINGS

This section starts by presenting the results of the three-dimensional synthetic data evaluation (Figure 1) applied to the three selected datasets (Table 1). These results provide the grounds for answering our first two research questions (RQ1 and RQ2). Then, we present results relevant to RQ3, namely, how to select appropriate synthetic data generation strategies for the use in different LA scenarios. Lastly, we discuss the limitations of the study.

### 4.1 Comprehensive evaluation of synthetic tabular data

**Resemblance.** Table 2 shows the average Jensen-Shannon Divergence (JSD), average Wasserstein Distance (WD) and Difference in pairwise correlation between real and synthetic data. The table demonstrates that the GC algorithm has the lowest while GM has the highest JSD values, meaning that GC performs the best and GM performs the worst in terms of resemblance metrics on categorical variables. This may be related to the fact that GM is a statistical distribution-based method, which is relatively simple and performs poorly on more complex data types and structures [28]. Additionally, GM has lowest WD, which indicates the smallest difference between the synthetic data generated by GM and the real data in terms of continuous variables. Furthermore, the difference in pairwise correlation is the smallest in case of GM generated data, meaning that such synthetic data captures the underlying relationships present in the real data relatively well. To sum up, to maximize synthetic data quality along the resemblance dimension, GM is recommended for datasets containing only continuous variables, whereas GC is recommended for datasets containing a high percentage of categorical variables, as GC performs well on both continuous and categorical variables. In addition, both GM and GC perform relatively well in pairwise correlation.

**Machine learning utility.** Table 3 shows, for each algorithm and each dataset, the computed measures of machine learning (ML) utility, namely differences between real and synthetic data in terms of Accuracy, F1 score, and ROC AUC. As the values presented in the table are computed by subtracting the ML utility of synthetic data from the utility values of real data, larger values in Table 3 mean worse performance of the synthetic data compared to the real data. As shown in Table 3, the highest ML utility difference is on the dataset B regardless of the algorithm used, which may be due to the fact that it is the only dataset that contains categorical data. Complex data types may pose more challenges. Comparing the three different algorithms for synthetic tabular data generation, GM has relatively superior Accuracy, F1 score, and ROC AUC on all three datasets, suggesting that in terms of ML utility, GM performs the best, followed by GC, and finally CTGAN. Combined with the results for the resemblance dimension (Table 2), this finding is consistent with the pattern that higher resemblance

---
[3] Hernadez et al. (2023)'s codes about MIA https://github.com/Vicomtech/STDG-evaluation-metrics

tends to imply higher utility [11].

Table 2. Average Jensen-Shannon Divergence (JSD), Average Wasserstein Distance (WD), and difference in pairwise correlation between real and synthetic data

| Algorithm | Dataset ID | JSD | WD | Correlation distance. diff |
|---|---|---|---|---|
| **GM** | A | 0.833 | 0.953 | 0.028 |
| | B | 0.684 | 0.085 | 2.573 |
| | C | 0.833 | 0.057 | 0.599 |
| **Average** | | 0.783 | **0.365** | **1.07** |
| **CTGAN** | A | 0.265 | 1.286 | 1.722 |
| | B | 0.02 | 0.18 | 8.783 |
| | C | 0.030 | 0.199 | 2.878 |
| **Average** | | 0.315 | 0.55 | 4.461 |
| **GC** | A | 0.081 | 1.857 | 0.105 |
| | B | 0.005 | 0.015 | 6.592 |
| | C | 0.049 | 0.206 | 2.537 |
| **Average** | | **0.045** | 0.693 | 3.078 |

**NA = Not Applicable**

Table 3. Difference in ML utility, measured through Accuracy, F1-score, and ROC AUC, between original and synthetic data

| Algorithm | Dataset ID | Accuracy Diff | F-1 score Diff | ROC AUC Diff |
|---|---|---|---|---|
| **GM** | A | 0.01 | 0.02 | 0.15 |
| | B | 0.14 | 0.15 | 0.08 |
| | C | 0.04 | 0.08 | 0.03 |
| **Average** | | 0.06 | **0.08** | **0.09** |
| **CTGAN** | A | 0.18 | 0.14 | 0.14 |
| | B | 0.36 | 0.42 | 0.34 |
| | C | 0.07 | 0.15 | 0.08 |
| **Average** | | 0.20 | 0.24 | 0.19 |
| **GC** | A | 0.01 | 0.02 | 0.16 |
| | B | 0.19 | 0.22 | 0.23 |
| | C | 0.07 | 0.12 | 0.01 |
| **Average** | | 0.09 | 0.12 | 0.13 |

**Privacy.** Table 4 shows the Distance to closest record (DCR) and Nearest neighbor distance ratio (NNDR) between original and synthetic data for privacy evaluation. A large difference between real and synthetic data indicates poor resemblance, as in such a case synthetic data fails to capture the overall structure of real data. However, if the value is very small, it indicates that there is a risk of exposing private information from the original data. When computing these metrics, we kept the $5th$ percentile of the metric value to provide a robust estimate of the privacy risk. Table 4 indicates that, on average, GM has the smallest value for DCR between real and synthetic data, which means it has the highest privacy risk. On the other hand, DCR and NNDR for CTGAN are the largest on all three datasets, meaning higher privacy measures. Lastly, in the case of the GM algorithm applied to dataset C, the DCR within synthetic data is significantly smaller than within real data, implying model collapse. Model collapse refers to the situation when generative models are replicating patterns that they have already seen and from which limited information can be extracted [40]. This means that in the case of multiple data generation using GM, the modeling errors increase in each iteration and consequently the misinterpretation of the data increases. That is, we cannot use GM repeatedly many times on the same dataset, as the study by [40] demonstrated that the synthetic data generated by multiple generators become increasingly similar to each other with little resemblance to the original data. Hence, this model collapse shows that it is feasible to use GM to generate one dataset once or few times, but it is not recommended to use GM repeatedly for the same dataset many times. This finding is important because for synthetic data generation, to avoid randomness, it is recommended to generate, for a given real dataset, multiple synthetic datasets using the same methods and choose the average value as the result. However, this feature of GM suggests that multiple rounds of data generation should be avoided as much as possible when using GM. It is difficult to detect this problem without the measurement of the distance-based privacy component. This also points to the importance of a comprehensive measurement of all synthetic data dimensions.

Table 4. Distance to closest record (DCR) and Nearest neighbor distance ratio (NNDR) between original and synthetic data.

| Algorithm | Dataset ID | DCR | | | NNDR | | |
|---|---|---|---|---|---|---|---|
| | | DCR between Real and Synthetic (5th perc) | DCR within Real (5th perc) | DCR within Synthetic (5th perc) | NNDR between Real and Synthetic (5th perc) | NNDR within Real (5th perc) | NNDR within Synthetic (5th perc) |
| **GM** | A | 0.188 | $0^*$ | 0.013 | 0.839 | NaN | 0.192 |
| | B | 3.746 | 2.897 | 3.885 | 0.792 | 0.43 | 0.777 |
| | C | 0.928 | 2.654 | 0.337 | 0.944 | 0.775 | 0.567 |
| **Average** | | 1.621 | 1.85 | 1.412 | 0.858 | 0.603 | 0.512 |
| **CTGAN** | A | 0.887 | 0 | 0.001 | 0.982 | NaN | 0.064 |
| | B | 6.708 | 2.897 | 7.294 | 0.924 | 0.43 | 0.896 |
| | C | 0.792 | 1.147 | 0.704 | 0.852 | 0.401 | 0.659 |
| **Average** | | **2.796** | **1.348** | **2.666** | **0.919** | **0.412** | **0.54** |
| **GC** | A | 0.622 | 0 | 0 | 1 | NaN | NaN |
| | B | 4.176 | 2.897 | 4.166 | 0.783 | 0.43 | 0.795 |
| | C | 0.484 | 1.148 | 0.091 | 0.904 | 0.493 | 0.355 |
| **Average** | | 1.76 | 1.348 | 1.419 | 0.896 | 0.461 | 0.575 |

\* DCR within real dataset is 0 means that a data point in the real dataset is very close to another data point in that dataset. This also apply to why NNDR within real is NaN, as it was caused by DCR within real is 0

Regarding the re-identification risk assessment, in this context, attack-based method, Table 5 presents the precision and accuracy of the specific MIA. Accuracy is the percentage of correct predictions made by the attacker, whereas precision denotes the proportion of records used for training the synthetic data generation approach identified by the attacker [20]. Since the approach of [20] simulates this attack by using Hamming distance thus assessing the success rate of the attacker, the thresholds 0.4, 0.3, 0.2 and 0.1 in the Table 5 column represent different Hamming distances used to measure the similarity between the records. Based on the different thresholds (0.4, 0.3, 0.2, and 0.1), the attacker determines whether there are sufficiently similar records of synthetic data to determine whether the student record $r$ is used to generate synthetic data. If directly interpret result, take GM, dataset A and thresholds 0.4's values as an example. The accuracy of the attacker's inferences is relatively high (Accuracy of 0.79), but the quality of the inferences is poor (Precision of 0). Furthermore, to better understand the results of MIA accuracy and precision, we follow the interpretation method of [20]: both accuracy and precision values below 0.5 indicate the best privacy preservation (assigned a value of 3), while a value of 0.5-0.8 for either accuracy or precision indicates moderate privacy preservation (assigned a value of 2); any value above 0.8 is considered poor privacy preservation (assigned a value of 1). The resulting score, rounded to the nearest integer, gives a value between 1 and 3 that indicates the MIA score of synthetic data generators. It can be seen that CTGAN and GC are all 2, which indicates moderate privacy, GM has the lowest value 1, which indicates poor privacy preservation. However, combining the performance results of distance-based methods and attack-based methods, CTGAN performs better than GC, whereas GM is the worst in terms of privacy.

Table 5. Membership interface attack (MIA) of the datasets used and their precision and accuracy.

| Algorit-hm | Dataset ID | Threshold 0.4 | | Threshold 0.3 | | Threshold 0.2 | | Threshold 0.1 | | |
|---|---|---|---|---|---|---|---|---|---|---|
| | | Precision | Accuracy | Precision | Accuracy | Precision | Accuracy | Precision | Accuracy | Evaluation |
| **GM** | A | 0 | 0.79 | 0 | 0.79 | 0 | 0.79 | 0 | 0.79 | 2 |
| | B | 0.24 | 0.24 | 0.24 | 0.24 | 0.82 | 0.76 | 0 | 0.76 | 1 |
| | C | 0 | 0.8 | 0 | 0.8 | 0 | 0.8 | 0 | 0.8 | 1 |
| **Average** | | 0.08 | 0.61 | .0.08 | 0.61 | 0.27 | 0.78 | 0 | 0.78 | 1 |
| **CTGAN** | A | 0 | 0.79 | 0 | 0.79 | 0 | 0.79 | 0 | 0.79 | 2 |
| | B | 0 | 0.79 | 0 | 0.79 | 0 | 0.79 | 0 | 0.79 | 2 |
| | C | 0 | 0.74 | 0 | 0.74 | 0 | 0.74 | 0 | 0.74 | 2 |
| **Average** | | **0** | **0.77** | **0** | **0.77** | **0** | **0.77** | **0** | **0.77** | **2** |
| **GC** | A | 0.21 | 0.21 | 0.2 | 0.22 | 0.2 | 0.22 | 0.2 | 0.35 | 3 |
| | B | 0.18 | 0.18 | 0.18 | 0.18 | 0.18 | 0.18 | 0.17 | 0.76 | 3 |
| | C | 0 | 0.83 | 0 | 0.83 | 0 | 0.83 | 0 | 0.83 | 1 |
| **Average** | | **0.13** | **0.41** | **0.13** | **0.41** | **0.13** | **0.41** | **0.12** | **0.69** | **2** |

.

## 4.2 Customized data generation for different LA scenarios

In response to RQ3, this section examines three real-world LA scenarios of synthetic data use with varying requirements in terms of the three dimensions of synthetic data quality, namely resemblance, ML utility, and privacy (Figure 2).

Figure 2. Recommendation of synthetic data generation approach in distinct LA scenarios.

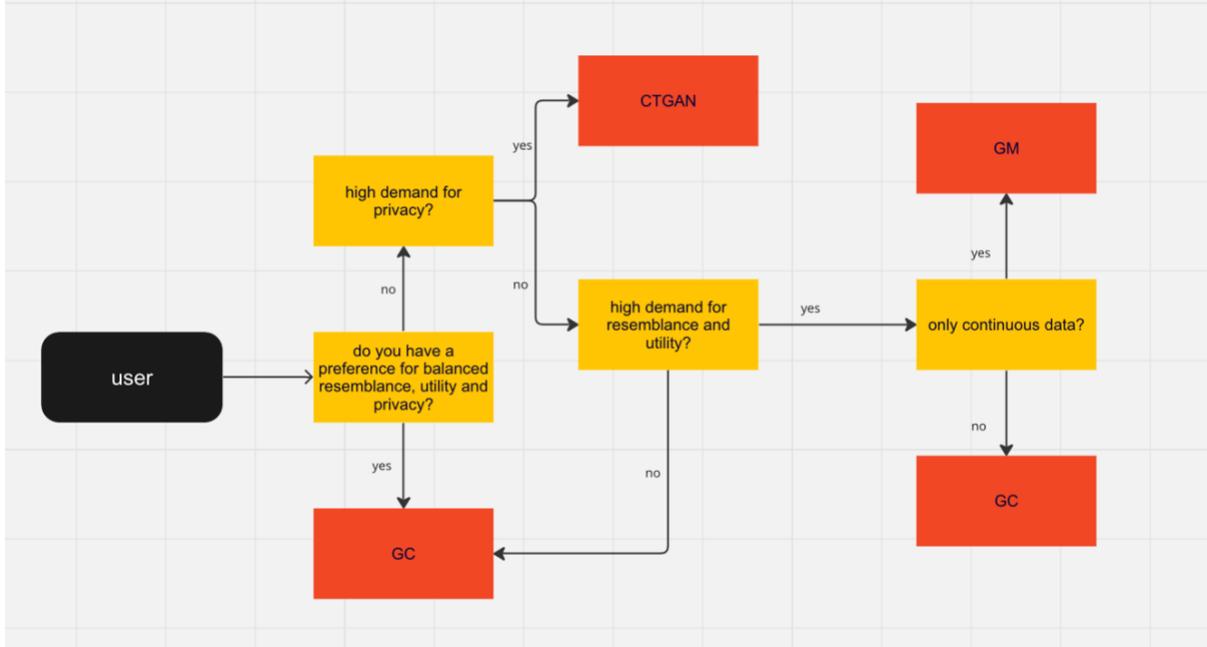

**Balanced preference scenario**: In this scenario, Alice, a LA researcher, faces data availability issues for a longitudinal study on predicting at-risk students due to discrepancies in consent at different time points (A and B). To address this, she opts for synthetic data to enrich her dataset. Alice wants the synthetic data to be similar to the real data and to have the ML utility similar to the real data. Furthermore, the use of synthetic data has to adhere to the privacy requirements because the students did not provide all the permissions on the real data. In such a situation, Alice can benefit from our comprehensive evaluation of synthetic data generation methods as it would allow her to choose the method that guarantees a balanced performance in terms of similarity, ML utility, and privacy. In particular, among the methods we examined in the current study, GC could offer such performance.

**High demand for utility and resemblance, low demand for privacy scenario**: Internal audit is required in evaluating and ensuring the quality of educational institutions in many places [13]. As the function of internal audit involves risk assessment, control testing, reporting and visualization [52], it has a high demand for data utility and not much demand for privacy. In such a case, we recommend the use of GM, which performs well in terms of resemblance and utility, if the audited data is completely continuous, and GC if it involves categorical data.

**High demand for privacy, low demand for utility scenario**: Such a scenario may occur when an educational institution needs to share a sensitive dataset with a third party in another country or region. According to the rules in many parts of the world, sensitive data must not be shared outside of the country/region, unless privacy and data protection is ensured.

Accordingly, the educational institution wants to go beyond the de-identification of the dataset and to generate highly private synthetic data to be shared with the third-party outside of the country. In such a scenario where privacy is highly emphasized, we recommend the use of CTGAN, which has a stable and good performance in terms of privacy.

**4.3 Study limitations**

The limitations of this paper are threefold. First, in order to exclude the effect of randomness and bias, [58] and [10] repeated generation of synthetic data and averaged the evaluation measures computed in individual iterations of data synthesis. Due to time and computation constraints, we did not use this method, but intend to do so to improve the robustness of the results. Second, there are a variety of measures that can be used for each of the examined synthetic data dimensions. While we aimed for a selection representative of this variety, the selection of some other measures might have led to different conclusions. Third, while this study has explored and presented scenarios addressing resemblance, utility, and privacy, we acknowledge that other unique cases and alternative solutions may exist beyond the scope of our synthetic data approach, warranting further consideration in future research. Finally, the data type and size of examined datasets is limited. There is a lot of time-series data in LA; as this type of data requires a different set of methods for synthetic data generation, it remained out of scope of the current study. Regarding the size and number of datasets, the current study did not examine truly "big" datasets (e.g., datasets with 10K or 100K records), as well as more than three datasets. A study of synthetic time-series data and big data are part of the plans for future work.

**5 DISCUSSION AND CONCLUSIONS**

The current study extends the boundaries of existing research on synthetic data within the LA field, paving a path forward by proposing a comprehensive evaluation of synthetic tabular data in terms of resemblance, ML utility, and privacy (RQ1). To assess resemblance, we include JSD and WD, which compare datasets in terms of probability distributions, and pairwise correlation, which measures the correlation between pairs of variables. Regarding ML utility, we compare the performance of ML predictive models, built on the original and synthetic data, using three ML methods commonly used in LA. In terms of privacy, we use representatives of both distance-based methods and attack-based methods to measure it. The three-dimensional evaluation can help users evaluate synthetic data comprehensively and use synthetic data generation methods that offer the best performance along the relevant dimension(s). Simply put, such a comprehensive evaluation allows for better informed decisions regarding synthetic data use, thus contributing to higher acceptance of synthetic data in the LA field.

Through the three-dimensional evaluation, our results confirm a trade-off between privacy and ML utility, e.g., GM performs better in utility but not as good in privacy compared to the other two methods. But we show that it is possible to use synthetic data to enhance the privacy of a dataset while maintaining similar machine learning performance as the real dataset (RQ2). This finding could help address the privacy and data utility concerns of educational institutions and LA researchers, making them more willing to release open access datasets, i.e., synthetic versions of the real confidential data. The release of more accessible and usable open datasets can address the increasing needs for large volume data sets driven by ongoing development of artificial intelligence [1] and ultimately spur the growth of the LA field.

Since different synthetic data generators have their own strengths and weaknesses in each of the three dimensions of resemblance, utility, and privacy, we hypothesized three different LA scenarios and gave recommendations for synthetic data generation based on the scenarios' needs (RQ3). The flexibility of synthetic data makes it applicable to the varying needs of different LA scenarios. Moreover, synthetic data also helps to combat possible data misuse privacy issues in LA. This is because synthetic data aims to mimic the structure of real data rather than the characteristics of individuals [10]. This avoids, to some extent, the issues arising from data being used for purposes other than those for which consent was given.

To facilitate the use of synthetic data in LA, we currently work on a platform that enables easy-to-use and easy-to-scale synthetic data generation called lasd.ai. Furthermore, our lasd.ai platform, responding to calls from previous researchers [5, 43], works to build an easy-to-use, easy-to-scalable synthetic data generator.

Finally, synthetic data, despite its recognized advantages, carries some risks of being used maliciously or as a means to bypass data protection legislation, if not used properly [3]. Therefore, for the future steps, while we explore the use of synthetic data in the LA domain for multivariate data types (e.g., time-series data), we should also explore how to prevent the misuse of synthetic data before malicious use of synthetic data occurs.